\documentstyle[12pt,amssymb,eqsecnum,aps]{revtex}
\newfont{\DamirFont}{arxi}
\newfont{\blackboard}{msbm10}
\newcommand{\Z}{\mbox{\blackboard\symbol{"5A}}}

\def\equals{\mathop{=}}
\newcommand{\io}{[\hspace{-1pt}[}
\newcommand{\ic}{]\hspace{-1pt}]}
\newcommand{\fo}{\{\!\mid\!}
\newcommand{\fc}{\!\mid\!\}}

\newcommand{\sgn}{{\rm sgn}}
\newcommand{\Det}{{\rm Det}}
\newcommand{\BS}{{\rm BS}}

\newcommand{\E}{{\cal E}}
\renewcommand{\Re}{{\rm Re\,}}
\newcommand{\REG}{{\rm REG}}
\newcommand{\IRREG}{{\rm IRREG}}

\newcommand{\ren}{{\rm ren}}
\newcommand{\eff}{{\rm eff}}
\newcommand{\tr}{{\rm tr}}

\renewcommand{\tanh}{{\rm tanh}}
\renewcommand{\arctan}{{\rm arctan}}

\newcommand{\tE}{\tilde{E}}
\newcommand{\tLp}{\tilde{L}_{(+)}}
\newcommand{\tLm}{\tilde{L}_{(-)}}

\def\be{\begin{equation}}
\def\ee{\end{equation}}
\def\bea{\begin{eqnarray}}
\def\eea{\end{eqnarray}}
\def\ba{\begin{array}}
\def\ea{\end{array}}
\def\V{{\bf V}}

\def\vac{{\rm vac}}
\def\x{{\bf x}}
\def\F{\Phi^{(0)}}

\newcommand{\bpar}{\mbox{\boldmath $\partial$}}
\newcommand{\ab}{\mbox{\boldmath $\alpha$}}
\newcommand{\gb}{\mbox{\boldmath $\gamma$}}

\def\si{\mathop{\displaystyle\sum\mkern-25mu\int\,}}

\newcommand{\ds}{\displaystyle}

\begin{document}
\title{\bf Self-adjointness of the twodimensional massless
Dirac Hamiltonian and vacuum energy density in
the background of\\ 
a singular magnetic vortex}

\author{{\large\bf Yurii A. Sitenko\thanks{Electronic address: yusitenko@bitp.kiev.ua}}\\
Bogolyubov Institute for Theoretical Physics, National Academy of
Sciences,\\ 03143 Kyiv, Ukraine\\}

\maketitle

\vspace{0.5cm}
({\large\bf Ukrainian Journal of Physics, 45, no.4/5, 569-578 (2000)})

\vspace{0.5cm}

\begin{abstract}
A massless spinor field is quantized in the background of a singular
static magnetic vortex in 2+1-dimensional space-time. The method of
self-adjoint extensions is employed to define the most general set
of physically acceptable boundary conditions at the location of the
vortex. Under these conditions, the vacuum energy density and effective 
potential in the vortex background are determined.
\end{abstract}

\section{Introduction}
Singular (or contact or zero-range) interaction potentials were
introduced in  quantum mechanics more than sixty years ago \cite{Bet,Tho,Fer}.
Since that time the attitude of physicists and mathematicians to this subject
was varying, starting from "it is impossible", then to "it is evident", and
finally arriving at "it is interesting" (for a review see monograph
\cite{Alb}). A mathematically consistent and rigorous treatment of the subject
was developed \cite{Ber}, basing on the notion of self-adjoint extension
of a Hermitian (symmetric) operator.

 Singular interaction is involved in quantum field theory when, for example,
a spinor field is quantized in the background of a pointlike magnetic monopole
in threedimensional space or a pointlike magnetic vortex in twodimensional
space. In these cases the Dirac Hamiltonian, in contrast to the Schrodinger one,
is free from an explicit $\delta$-function singularity;
nonetheless the problem of self-adjoint extension of both Dirac and Schrodinger
operators arises, albeit for different reasons (see, for example \cite{Gacki}).
A distinguishing feature is that a
solution to the Dirac equation, unlike that to the Schrodinger one, cannot
obey a condition of regularity at the singularity point. It is
necessary then to define a boundary condition at this point, and the
least restrictive, but still physically acceptable, condition is such that
guarantees self-adjointness of the Dirac Hamiltonian. Thus, effects of
polarization of the fermionic vacuum in a singular background
(such as a pointlike monopole or a pointlike vortex) appear to
depend on the choice of the boundary condition at the singularity
point, and a set of permissible boundary conditions is labelled, most
generally, by the values of self-adjoint extension parameters.
In contrast to the Schrodinger case, the extension in the Dirac case does not
reflect additional types of interaction but represents complementary
information that must be specified when describing the physical attributes
of the already posited singular background configuration.

As a consequence, the fermionic vacuum under the influence of a
singular background can acquire rather unusual properties: leakage of
quantum numbers from the singularity point  occurs. While in the case
of a monopole there
is leakage of charge to the vacuum, which results in the monopole
becoming the dyon violating the Dirac quantization condition and CP
symmetry \cite{Gol,Cal,Wit,Gro,Yam}, in the case
of a vortex (the Ehrenberg-Siday-Aharonov-Bohm potential \cite{Ehre,Aha})
the situation is much
more complicated, since there is leakage of both charge and other
quantum numbers to the vacuum.
Apparently, this is due to a nontrivial topology of the base space in the
latter case: $\pi_1=0$ in the case of a space with a deleted point, and
$\pi_1=\Z$ in the case of a space with
a deleted line (or a plane with a deleted point); here $\pi_1$ is the
first homotopy group and $\Z$ is a set of integer numbers.
For a particular choice of the boundary
condition at the location of a singular vortex it has been shown that charge
\cite{Sit88,Sit90}, current \cite{Gor} and angular momentum
\cite{SitR96} are induced in the vacuum. The induced vacuum quantum
numbers under general boundary conditions which are  compatible with
self-adjointness have been considered in Refs.\cite{Sit96,Sit97,SitR97,Sit99}.

A pointlike static magnetic vortex (the Ehrenberg-Siday-Aharonov-Bohm
configuration) in $2+1$ dimensional space-time has the
form 
$$V^1(\x)=-\Phi^{(0)}{x^2\over(x^1)^2+(x^2)^2}, \quad V^2(\x)=\F
{x^1\over(x^1)^2+(x^2)^2}, \eqno(1.1)$$
$$\bpar\times\V(\x)=2\pi\F\delta(\x),
\eqno(1.2) $$
where $\F$ is the vortex flux in $2\pi$ units, i.e.
in the London ($2\pi \hbar c e^{-1}$) units, since we use
conventional units $\hbar=c=1$ and coupling constant $e$ is included into
vector potential $\V(\x)$.
The wave function on the plane ($x^1$, $x^2$) with punctured singular
point $x^1=x^2=0$
obeys the most general condition (see \cite{Sit96} for more details)
$$<r,\varphi+2\pi|=e^{i2\pi\Upsilon}<r,\varphi| \,  , \eqno(1.3)$$
where $r=\sqrt{(x^1)^2+(x^2)^2}$ and $\varphi=\arctan(x^2/x^1)$ are the
polar coordinates, and $\Upsilon$ is a continuous real parameter which
is varied in the range $0\leq\Upsilon<1$. It can be shown (see, for
example, \cite{SitR96,Sit96}) that $\Upsilon$ as well as $\F$ is
changed under singular gauge transformations, whereas difference
$\F-\Upsilon$ remains invariant. Thus, physically sensible quantities
are to depend on the gauge invariant combination $\F-\Upsilon$ which
will be for brevity denoted as the reduced vortex flux in the following.

Thus far the effects of polarization of the massive fermionic vacuum
have been studied. In the present paper we find the energy density which is 
induced by a singular vortex in the massless fermionic vacuum.
In the next section, the most general set of physically acceptable boundary
conditions at the singularity point $\x=0$ is defined. In Section III we show 
how the problem of both ultraviolet and
infrared divergences in vacuum characteristics is solved with the help of zeta
function regularization. This allows us to get immediately
in Section IV the vacuum energy density; also here the effective action and the
effective potential are considered. We summarize
results and discuss their consequences in Section V. The method of self-adjoint 
extension is employed to get the boundary condition at the singularity point 
in Appendix.

\section{Quantization of a Spinor Field and the Boundary Condition at the
Location of a Vortex}

The operator of the second-quantized spinor field is presented in the
form
$$
\Psi(\x,t)=\si_{E_\lambda>0}\, e^{-iE_\lambda
t}<\x|\lambda>a_\lambda+\si_{E_\lambda<0}\, e^{-iE_\lambda
t}<\x|\lambda>b_\lambda^+, \eqno(2.1)
$$
where $a_\lambda^+$ and $a_\lambda$ ($b_\lambda^+$ and $b_\lambda$) are
the spinor particle (antiparticle) creation and annihilation operators
satisfying anticommutation relations
$$
[a_\lambda,a_{\lambda'}^+]_+=[b_\lambda,b_{\lambda'}^+]_+=<\lambda|\lambda'>,
\eqno(2.2)
$$
and $<\x|\lambda>$ is the solution to the stationary Dirac equation
$$ H<\x|\lambda>=E_\lambda<\x|\lambda>, \eqno(2.3)$$
$H$ is the Dirac Hamiltonian, $\lambda$ is the set of parameters
(quantum numbers) specifying a state, $E_\lambda$ is the energy of a
state; symbol $\si$ means the summation over discrete and the
integration (with a certain measure) over continuous values of
$\lambda$. The ground state $|\vac>$ is defined conventionally by
equality
$$ a_\lambda|\vac>=b_\lambda|\vac>=0. \eqno(2.4) $$
In the case of quantization of a massless spinor field in the
background of static vector field $\V(\x)$, the Dirac Hamiltonian
takes the form
$$ H=-i\ab[\bpar-i\V(\x)], \eqno(2.5)$$
where
$$\ab=\gamma^0\gb, \qquad \beta=\gamma^0, \eqno(2.6)$$
$\gamma^0$ and $\gb$ are the Dirac $\gamma$ matrices. In the
2+1-dimensional space-time $(\x,t)=(x^1,x^2,t)$ the Clifford algebra
has two inequivalent irreducible representations which can be differed
in the following way:
$$i\gamma^0\gamma^1\gamma^2=s, \qquad s=\pm1.
\eqno(2.7)$$
Choosing the $\gamma^0$ matrix in the diagonal form
$$\gamma^0=\sigma_3 , \eqno(2.8)$$
one gets
$$\gamma^1=e^{{i\over2}\sigma_3\chi_s}i\sigma_1e^{-{i\over2}\sigma_3\chi_s},
\quad \gamma^2=
e^{{i\over2}\sigma_3\chi_s}is\sigma_2e^{-{i\over2}\sigma_3\chi_s},
\eqno(2.9)$$
where $\sigma_1,\sigma_2$ and $\sigma_3$ are the Pauli matrices,
and $\chi_1$ and
$\chi_{-1}$ are the parameters that are varied in the interval
$0\leq\chi_s<2\pi$ to go over to the equivalent representations.

A solution to the Dirac equation (2.3) with Hamiltonian (2.5) in
background (1.1), that obeys condition (1.3), can be
presented as
$$<\x|E,n>=\left(\ba{l}
f_n(r,E)e^{i(n+\Upsilon)\varphi}\\[0.2cm]
g_n(r,E)e^{i(n+\Upsilon+s)\varphi}\\ \ea \right), \quad n\in\Z,
\eqno(2.10)$$
where the column of radial functions
$\left(\ba{l}
f_n\\
g_n\\ \ea \right)$
satisfies the equation
$$h_n\left(\ba{l}
f_n\\[0.2cm]
g_n\\ \ea \right)=
E\left(\ba{l}
f_n\\[0.2cm]
g_n\\ \ea \right),
\eqno(2.11)$$
and
$$
h_n=
\left(
\ba{l}
\quad \quad \quad \quad \quad \quad \quad 0\\[0.2cm]
e^{-i\chi_s}[-\partial_r+s(n-\F+\Upsilon)r^{-1}] \\
\ea
\ba{l}
e^{i\chi_s}[\partial_r+s(n-\F+\Upsilon+s)r^{-1}]\\[0.2cm]
\quad \quad \quad \quad \quad \quad \quad 0 \\
\ea
\right)
\eqno(2.12)$$
is the partial Dirac Hamiltonian.
When reduced vortex flux $\F-\Upsilon$ is integer, the requirement of
square integrability for wave function (2.10) at $r\rightarrow 0$
provides its regularity, rendering the partial Dirac
Hamiltonian $h_n$ for every value of $n$ to be essentially self-adjoint.
When $\F-\Upsilon$ is fractional, the same is valid only for $n\neq
n_0$, where
$$n_0=\io\F-\Upsilon\ic+{1\over2}-{1\over2}s, \eqno(2.13)$$
$\io u\ic$ is the integer part of a quantity $u$ (i.e., the greatest
integer that is less than or equal to $u$). For $n=n_0$, each of the
two linearly independent solutions to Eq.(2.11) meets the
requirement of square integrability at $r\rightarrow 0$. Any particular
solution in this
case is characterized by at least one (at most both) of the radial
functions being divergent as $r^{-p}$ ($p<1$) at $r\rightarrow 0$. If
one of the two linearly independent solutions is chosen to have a
regular upper and an irregular lower component, then the other one has
a regular lower and an irregular upper component. Therefore,
in contrast to operator $h_n$ ($n\neq n_0$), operator
$h_{n_0}$ is not essentially self-adjoint
\footnote{A corollary of the
theorem proven in Ref.\cite{Wei} states that, for the partial Dirac
Hamiltonian to be essentially self-adjoint, it is necessary and
sufficient that a non-square-integrable (at $r\rightarrow 0$) solution exist.}.
The Weyl - von Neumann theory of self-adjoint operators (see, e.g.,
Refs.\cite{Alb,Akhie}) has to be employed in order to consider the
possibility of a self-adjoint extension in the case of $n=n_0$. It is shown
in Appendix that the self-adjoint extension exists indeed and is
parametrized by one continuous real variable denoted in the following
by $\Theta$. Thus operator $h_{n_0}$
is defined on the domain of functions obeying the condition

$$\cos\bigl(s{\Theta\over2}+{\pi\over4}\bigr)\lim_{r\rightarrow 0}(\mu
r)^Ff_{n_0}=-e^{i\chi_s}\sin\bigl(s{\Theta\over2}+{\pi\over4}\bigr)
\lim_{r\rightarrow 0}(\mu r)^{1-F}g_{n_0}, \eqno(2.14)$$
where $\mu>0$ is the parameter of the dimension of inverse length and
$$
F=s\fo\F-\Upsilon\fc+{1\over2}-{1\over2}s,
\eqno(2.15)$$
$\fo u\fc=u-\io u\ic$ is the fractional part of a quantity $u$,
$0\leq\fo u\fc<1$; note here that Eq.(2.14) implies that $0<F<1$,
since in the case of $F={1\over2}-{1\over2}s$ both $f_{n_0}$ and
$g_{n_0}$ obey the condition of regularity at $r\rightarrow 0$. Note
also that Eq.(2.14) is periodic in $\Theta$ with period $2\pi$; therefore,
without a loss of generality, all permissible values of $\Theta$ will be
restricted in the following to range $-\pi\leq\Theta\leq\pi$.

\section{Zeta Function}

In the second-quantized theory the operator of energy is defined as

$$
\hat{\E}=\int d^2x\,{1\over2}\bigl[\Psi^+(\x,t), H\Psi(\x,t)\bigr]_- =\si
(E_\lambda a_\lambda^+a_\lambda -E_\lambda b_\lambda^+ b_\lambda
-{1\over2}|E_\lambda|), \eqno(3.1)
$$
thus the vacuum expectation value of the energy takes the form

$$
\E\equiv <\vac|\,\hat{\E}|\vac>=-{1\over2} \si|E_\lambda|=- {1\over2}\int
d^2x\,\tr<\x|\,|H|\,|\x>. \eqno(3.2)$$
The latter expression is ill-defined due to divergences of various
kinds. First, there is a bulk divergence resulting from the integration
over the infinite twodimensional space. But, even if one considers the
vacuum energy density,

$$\E_\x=-{1\over2}\tr<\x|\,|H|\,|\x>, \eqno(3.3)$$
still it remains to be divergent. There is a divergence at large values
of momentum of integration, $k\rightarrow \infty$. To tame this
divergence, let us introduce the zeta function density

$$\zeta_\x(z)=\tr<\x|\,|H|^{-2z}|\x>, \eqno(3.4)
$$
which is ultraviolet convergent at sufficiently large values of $\Re
z$. However, exactly at these values of $\Re z$ the integral
corresponding to Eq.(3.4) is divergent in the infrared region, as
$k\rightarrow 0$. To regularize this last divergence, let us introduce
fermion mass $m$, modifying definition (3.4):
$$
\zeta_\x(z|m)=\tr<\x|\,|\tilde{H}|^{-2z}|\x>, \eqno(3.5)
$$
where
$$
\tilde{H}=-i\ab[\bpar-i\V(\x)] +\beta m, \eqno(3.6)
$$
and it is implied that the complete set of solutions to the equation
$$
\tilde{H}<\x|\lambda>=\tilde{E}_\lambda<\x|\lambda>, \eqno(3.7)
$$
instead of those to Eq.(2.3), is used.

In the background of a singular magnetic vortex (1.1) -- (1.2) the
radial functions of the solutions to Eq.(3.7) take the form:

$$
\left(\ba{c}
\tilde{f}_n\\
\tilde{g}_n\\ \ea
\right) =
\left(\ba{c}
\ds{\sqrt{1+m\tilde{E}^{-1}}J_{l-F}(kr)e^{i\chi_s}}\\[0.2cm]
\ds{\sgn(\tilde{E})\sqrt{1-m\tilde{E}^{-1}}J_{l+1-F}(kr)}\\
\ea\right), \qquad l=s(n-n_0)>0, \eqno(3.8)
$$

$$
\left(\ba{c}
\tilde{f}_n\\
\tilde{g}_n\\ \ea
\right) =
{1\over2\sqrt{\pi}}
\left(\ba{c}
\ds{\sqrt{1+m\tilde{E}^{-1}}J_{l'+F}(kr)e^{i\chi_s}}\\[0.2cm]
\ds{-\sgn(\tilde{E})\sqrt{1-m\tilde{E}^{-1}}J_{l'-1+F}(kr)}\\
\ea\right), \qquad l'=s(n_0-n)>0, \eqno(3.9)
$$

$$
\left(\ba{c}
\tilde{f}_{n_0}^{(C)}\\
\tilde{g}_{n_0}^{(C)}\\ \ea
\right) =
{1\over 2\sqrt{\pi[1+\sin(2\tilde{\nu}_{\tilde{E}})\cos(F\pi)}}\times $$
$$\times
\left(\ba{c}
\ds{\sqrt{1+m\tilde{E}^{-1}}[\sin(\tilde{\nu}_{\tilde{E}})J_{-F}(kr)+
\cos(\tilde{\nu}_{\tilde{E}})J_F(kr)]e^{i\chi_s}}\\[0.2cm]
\ds{\sgn(\tilde{E})\sqrt{1-m\tilde{E}^{-1}}[\sin(\tilde{\nu}_{\tilde{E}})
J_{1-F}(kr)-\cos(\tilde{\nu}_{\tilde{E}})J_{-1+F}(kr)]}\\
\ea\right), \eqno(3.10)
$$
where $k=\sqrt{\tilde{E}^2-m^2}$, $J_{\rho}(u)$ is the Bessel function of order
$\rho$ and
$$
\tan(\tilde{\nu}_{\tilde{E}})=\sgn(\tilde{E})\sqrt{{1-m\tilde{E}^{-1}\over
1+m\tE^{-1}}}\left({k\over 2\mu}\right)^{2F-1}{\Gamma(1-F)\over\Gamma(F)}
\tan\left(s{\Theta\over2}+{\pi\over4}\right) , \eqno(3.11)
$$
$\Gamma(u)$ is the Euler gamma function; note that the radial functions of
irregular solution (3.10) satisfy condition (2.14) (see Appendix). 
Note also that Eqs.(3.8) -- (3.10) correspond to the continuum,
$|\tE|>|m|$ \footnote{In a 2+1-, as well as in any odd-, dimensional
space-time mass parameter $m$ in Eq.(3.6) can take both positive
and negative values.}. In addition to them, in the case of
$$
\sgn(m)\cos \Theta<0 ,\eqno(3.12)
$$
an irregular solution corresponding to the bound state appears. Its
radial functions are
$$
\left(\ba{c}
\tilde{f}_{n_0}^{(\BS)}\\
\tilde{g}_{n_0}^{(\BS)}\\ \ea
\right) =
{\kappa\over\pi} \sqrt{{\sin(F\pi)\over 1+(2F-1)m^{-1}E_\BS}}
\left(\ba{c}
\ds{\sqrt{1+m^{-1}E_\BS}K_F(\kappa
r)e^{i\chi_s}}\\[0.2cm]
\ds{\sgn(m)\sqrt{1-m^{-1}E_\BS}
K_{1-F}(\kappa r)]}\\
\ea\right), \eqno(3.13)
$$
where $\kappa=\sqrt{m^2-E_\BS^2}$, $K_{\rho}(w)$ is the Macdonald 
function of order $\rho$ and the bound state energy
$\tE=E_\BS$ $(|E_\BS|<|m|)$ is determined implicitly by the equation
$$
{(1+m^{-1}E_\BS)^{1-F}\over (1-m^{-1}E_\BS)^F}
=-\sgn(m)\left({|m|\over 2\mu}\right)^{2F-1} {\Gamma(1-F)\over\Gamma(F)}
\tan\left(s{\Theta\over2}+{\pi\over4}\right). \eqno(3.14)
$$

Regular solutions (3.8) and (3.9) yield the following contribution
to zeta function density (3.5):
$$
[\zeta_\x(z|m)]_\REG={1\over4\pi} \int\limits_0^\infty
dk\,k|\tE|^{-2z}\sum_{\sgn(\tE)} \bigl\{ \sum_{l=1}^\infty
\bigl[(1+m\tE^{-1})J_{l-F}^2(kr)
+(1-m\tE^{-1})J_{l+1-F}^2(kr)\bigr]+$$
$$+\sum_{l'=1}^\infty
\bigl[(1+m\tE^{-1})J_{l'+F}^2(kr)+
(1-m\tE^{-1})J_{l'-1+F}(kr)\bigr]\bigr\}. \eqno(3.15)
$$
Summing over the energy sign and over $l$ and $l'$, we get the
expression
$$
[\zeta_\x(z|m)]_\REG ={1\over\pi} \int\limits_0^\infty dk\,k
|\tE|^{-2z} \int\limits_0^{kr}{dy\over y} \bigl[FJ_F^2(y)+(1-F)
J_{1-F}^2(y)\bigr], \eqno(3.16)
$$
which in the case of $\Re z>1$ is reduced to the form
$$
[\zeta_\x(z|m)]_\REG= {1\over2\pi(z-1)} \int\limits_0^\infty {dk\over
k} |\tE|^{2-2z}\bigl[ FJ_F^2(kr)+(1-F)J_{1-F}^2(kr)\bigr]. \eqno(3.17)
$$

Irregular solution (4.10) yields the following contribution to
Eq.(3.5):
$$
[\zeta_\x(z|m)]_\IRREG ={1\over4\pi} \int\limits_0^\infty
dk\,k|\tE|^{-1-2z}\bigl\{ A\mu^{1-2F}k^{2F}[\tLp-\tLm]J_{-F}^2(kr)+$$
$$+ A\mu^{1-2F}k^{-2(1-F)} [(m-|\tE|)^2\tLp-(m+|\tE|)^2\tLm]
J_{1-F}^2(kr)+$$
$$ +2[(m+|\tE|)\tLp-(m-|\tE|)\tLm] J_{-F}(kr)J_F(kr)+2[(m-|\tE|)\tLp-$$
$$ -(m+|\tE|)\tLm] J_{1-F}(kr)J_{-1+F}(kr)+ A^{-1}\mu^{2F-1}k^{-2F}
[(m+|\tE|)^2\tLp-$$
$$ -(m-|\tE|)^2\tLm] J_F^2(kr)+A^{-1}\mu^{2F-1}k^{2(1-F)} [\tLp-\tLm]
J_{-1+F}^2(kr)\bigr\}, \eqno(3.18)
$$
where summation over the energy sign has been performed and
$$
A=2^{1-2F}{\Gamma(1-F)\over\Gamma(F)}\tan\left(s{\Theta\over2}+
{\pi\over4}\right), \eqno(3.19)$$
$$
\tilde{L}_{(\pm)}=[A\mu^{1-2F}k^{-2(1-F)}(-m\pm|\tE|)+
2\cos(F\pi)+A^{-1}\mu^{2F-1}k^{-2F} (m\pm|\tE|)]^{-1}. \eqno(3.20)
$$

The contribution of bound state solution (3.13) to Eq.(3.5) is the
following:
$$
[\zeta_\x(z|m)]_\BS= {\sin(F\pi)\over\pi^2}\,
{\kappa^2|E_\BS|^{-2z}\over m+E_\BS(2F-1)} \bigl[(m+E_\BS)K_F^2(\kappa
r)+(m-E_\BS)K_{1-F}^2(\kappa r)\bigr]. \eqno(3.21)
$$

By deforming the contour of integration in the complex w-plane, Eq.(3.17) 
in the case of $1<\Re z<2$ is
transformed to the following expression
$$
[\zeta_\x(z|m)]_\REG ={|m|^{2(1-z)}\over 2\pi(z-1)} +{\sin(z\pi)\over
\pi^2(z-1)} r^{2(z-1)} \times $$
$$\times \int\limits_{|m|r}^\infty {dw\over
w}(w^2-m^2r^2)^{1-z} \bigl[ FI_F(w)K_F(w)+(1-F)I_{1-F}(w)K_{1-F}(w)\bigr],
\eqno(3.22)
$$
while Eq.(3.18) in the case of ${1\over2}<\Re z<1$ is transformed to
the following one
$$
[\zeta_\x(z|m)]_\IRREG ={\sin(z\pi)\over\pi^2} r^{2(z-1)}
\int\limits_{|m|r}^\infty dw\, w(w^2-m^2r^2)^{-z} \bigl[I_F(w)K_F(w)+
I_{1-F}(w)K_{1-F}(w)\bigr]+$$
$$+{2\sin(F\pi)\over\pi^3}
\sin(z\pi)r^{2(z-1)} \times $$
$$\times \int\limits_{|m|r}^\infty dw\,
w(w^2-m^2r^2)^{-z} { A\mu^{1-2F}({w\over r})^{2F}K_F^2(w)+A^{-1}\mu^{2F-1}
({w\over r})^{2(1-F)}K_{1-F}^2(w)\over  A\mu^{1-2F}({w\over
r})^{2F}+2m+A^{-1}\mu^{2F-1}({w\over r})^{2(1-F)}}-$$
$$-{\sin(F\pi)\over\pi^2}\, {\kappa^2|E_\BS|^{-2z}\over m+E_\BS(2F-1)}
\bigl[(m+E_\BS)K_F^2(\kappa r)+(m-E_\BS)K_{1-F}^2(\kappa r)\bigr];
\eqno(3.23)
$$
here $I_\rho(w)$ is the modified Bessel function of order $\rho$.
The integral in Eq.(3.22) can be analytically continued to domain
${1\over2}<\Re z<2$. In the case of ${1\over2}<\Re z<1$ this integral
is decomposed into two terms:
$$[\zeta_\x(z|m)]_\REG= {|m|^{2(1-z)}\over 2\pi(z-1)}-$$
$$-{\sin(z\pi)\over\pi^2} r^{2(z-1)} \int\limits_{|m|r}^\infty dw\,
w(w^2-m^2r^2)^{-z} \bigl[I_F(w)K_F(w)+I_{1-F}(w)K_{1-F}(w)\bigr]+$$
$$+{2\sin(F\pi)\over\pi^3} \, {\sin(z\pi)\over z-1} r^{2(z-1)}
\int\limits_{|m|r}^\infty dw(w^2-m^2r^2)^{1-z} K_F(w)K_{1-F}(w),
\eqno(3.24)
$$
the last of which can be analytically continued to domain $\Re
z<2$. Note also that the second integral in Eq.(3.23) can be
analytically continued to domain $\Re z<1$.

Summing Eqs.(3.21), (3.23) and (3.24), we get
$$
\zeta_\x(z|m)= {|m|^{2(1-z)}\over 2\pi(z-1)}+ {2\sin(F\pi)\over\pi^3}
\, {\sin(z\pi)\over z-1} r^{2(z-1)} \int\limits_{|m|r}^\infty
dw(w^2-m^2r^2)^{1-z} K_F(w)K_{1-F}(w)+$$
$$+ {2\sin(F\pi)\over\pi^3} \sin(z\pi)r^{2(z-1)}\times$$
$$\times
\int\limits_{|m|r}^\infty dw\, w(w^2-m^2r^2)^{-z}
 {A\mu^{1-2F}({w\over r})^{2F}K_F^2(w)+A^{-1}\mu^{2F-1}({w\over
r})^{2(1-F)} K_{1-F}^2(w)\over A\mu^{1-2F}({w\over
r})^{2F}+2m+A^{-1}\mu^{2F-1}({w\over r})^{2(1-F)}}, \eqno(3.25)
$$
i.e., the terms which are defined only in domain ${1\over2}<\Re
z<1$ are cancelled.

Note that the first term in Eq.(3.25) is identified with the zeta
function density in the noninteracting theory (i.e. in the absence of
any boundary condition and any background field):
$$
\zeta_\x^{(0)}(z|m)={|m|^{2(1-z)}\over 2\pi(z-1)}. \eqno(3.26)
$$
In the noninteracting theory all vacuum values are simply omitted due to
the prescription of normal ordering of the product of operators (see,
for example, \cite{Itzy}). Therefore, one has to subtract
$\zeta_\x^{(0)}$ from $\zeta_\x$ for the reasons of consistency. Doing
this and removing the infrared regulator mass, we obtain the
renormalized zeta function density
$$
\zeta_\x^\ren(z) \equiv \lim_{m\rightarrow 0}
\bigl[\zeta_\x(z|m)-\zeta_\x^{(0)}(z|m)\bigr]=$$
$$={\sin(F\pi)\over\pi^3} \sin(z\pi)r^{2(z-1)} \Bigg\{
{\sqrt{\pi}\over4}\, {\Gamma(1-z)\over \Gamma({3\over2}-z)}
\left[1-2{F(1-F)\over1-z}\right] \Gamma(F-z)\Gamma(1-F-z)+$$
$$+\int\limits_0^\infty dw\, w^{1-2z}\left[K_F^2(w)-K_{1-F}^2(w)\right]
\tanh\left[(2F-1)\ln\left({w\over \mu r}\right)+\ln A\right]\Bigg\}. \eqno(3.27)
$$

\section{Energy Density and Effective Potential}

Recalling the formal expressions for the vacuum energy and zeta
function densities, Eqs.(3.3) and (3.4), one can easily deduce that the
physical (renormalized) vacuum energy density is expressed through the
renormalized zeta function density at $z=-{1\over2}$:
$$
\E_\x^\ren=-{1\over2}\zeta_\x^\ren(-{1\over2}). \eqno(4.1)
$$
In the background of a singular magnetic vortex (1.1) -- (1.2), using
Eq.(3.27), we get the expression
$$
\E_\x^\ren={\sin(F\pi)\over2\pi r^3} \biggl\{
{{1\over2}-F\over6\cos(F\pi)}\left[{3\over4}-F(1-F)\right]+$$
$$+{1\over\pi^2}
\int\limits_0^\infty dw\, w^2
\bigl[K_F^2(w)-K_{1-F}^2(w)\bigr]\tanh\bigl[
(2F-1)\ln({w\over\mu r})+\ln A\bigr]\biggr\}. \eqno(4.2)
$$
At noninteger values of reduced vortex flux $\F-\Upsilon$ (i.e.
at $0<F<1$) vacuum energy density (4.2) is positive. At half-integer
values of the reduced vortex flux ($F={1\over2}$) we get
$$
\E_\x^\ren\big|_{F={1\over2}} ={1\over24\pi^2 r^3}. \eqno(4.3)
$$
In the case of $\cos\,\Theta=0$ we get
$$
\E_\x^\ren={\tan(F\pi)\over4\pi r^3} \bigl(F-{1\over2}\bigr)
\left[{1\over3}F(1-F)-{1\over4}\mp{1\over2}\bigl(F-{1\over2}\bigr)
\right], \qquad \Theta=\pm s{\pi\over2}. \eqno(4.4)
$$
If $\cos\,\Theta\neq0$, then at large distances from the vortex we get
$$
\E_\x^\ren \equals_{r\rightarrow \infty} {\tan(F\pi)\over4\pi
r^3} \bigl(F-{1\over2}\bigr)
\left[{1\over3}F(1-F)-{1\over4}+{1\over2}|F-{1\over2}|\right].
\eqno(4.5)
$$

Going over to imaginary time $t=-i\tau$,  let us consider the
effective action in 2+1-dimensional Euclidean space-time:
$$
S_{(2+1)}^\eff[\V(\x)] =-\ln\bigl\{ N^{-1}\int d\Psi\,d\Psi^+\exp[-\int
d\tau\,d^2x\,\Psi^+(-i\beta\partial_\tau-
i\beta \tilde{H})\Psi]\bigr\}=$$
$$=-\ln\Det\bigl[(-i\beta\partial_\tau-
i\beta\tilde{H})\tilde{m}^{-1}\bigr]; \eqno(4.6)
$$
here $N$ is a normalization factor, parameter $\tilde{m}$ is
inserted just for the dimension reasons, while fermion mass $m$ (see
Eq.(3.6)) is introduced in order to tame the infrared divergence. The
real part of the effective action is presented in the form \footnote{The
imaginary part of the effective action vanishes in the case of a static
background.}
$$
\Re S_{(2+1)}^\eff[\V(\x)]=-{1\over2}\int d\tau
d^2x\,\tr<\x,\tau|\ln[(-\partial_\tau^2+\tilde{H}^2)\tilde{m}^{-2}]|\x,\tau>.
\eqno(4.7)
$$
Let us define the zeta function density in threedimensional space
($x^1,x^2,\tau$):
$$
\zeta_{\x,\tau}(z|m)=\tr<\x,\tau|\,(-\partial_\tau^2 +
\tilde{H}^2)^{-z}|\x,\tau>. \eqno(4.8)
$$
Then an ultraviolet regularization of Eq.(4.7) can be achieved by
expressing its integrand through Eq.(4.8):
$$
-{1\over2}\tr<\x,\tau|\ln[(-\partial_\tau^2+\tilde{H}^2)\tilde{m}^{-2}]
|\x,\tau>= {1\over2}\bigl[{d\over
dz}\zeta_{\x,\tau}(z|m)\bigr]\big|_{z=0} +
{1\over2}\zeta_{\x,\tau}(0|m)\ln\tilde{m}^2. \eqno(4.9)
$$

In the background of a singular magnetic vortex (1.1) -- (1.2) we
get, similarly to Eq.(3.25), the following expression
$$
\zeta_{\x,\tau}(z|m)={|m|^{3-2z}\over4\pi^{3\over2}}\,
{\Gamma(z-{3\over2})\over\Gamma(z)} -$$
$$-{\sin(F\pi)\over\pi^{7\over2}}
\cos(z\pi){\Gamma(z-{3\over2})\over\Gamma(z)}r^{2z-3}
\int\limits_{|m|r}^\infty dw(w^2-m^2r^2)^{{3\over2}-z}
 K_F(w)K_{1-F}(w)-$$
$$-{\sin(F\pi)\over\pi^{7\over2}}\cos(z\pi)
{\Gamma(z-{1\over2})\over\Gamma(z)} r^{2z-3}\times$$
$$\times\int\limits_{|m|r}^\infty
dw\, w(w^2-m^2r^2)^{{1\over2}-z}
 {A\mu^{1-2F}({w\over r})^{2F}K_F^2(w)+A^{-1}\mu^{2F-1}({w\over
r})^{2(1-F)}K_{1-F}^2(w)\over A\mu^{1-2F}({w\over r})^{2F}+2m+A^{-1}
\mu^{2F-1}({w\over r})^{2(1-F)}}; \eqno(4.10)
$$
note that the first term in Eq.(4.10) corresponds to the case of the
noninteracting theory:
$$
\zeta_{\x,\tau}^{(0)}(z|m)={|m|^{3-2z}\over4\pi^{3\over2}}\,
{\Gamma(z-{3\over2})\over\Gamma(z)}. \eqno(4.11)
$$
Note also relation
$$
\zeta_{\x,\tau}(0|m)=0, \eqno(4.12)
$$
which ensures the independence of the effective action on
$\tilde{m}^2$; thus Eq.(4.9) takes the form
$$
-{1\over2}\tr\langle
\x,\tau|\ln(-\partial_\tau^2+\tilde{H}^2)\tilde{m}^{-2}|\x,\tau\rangle=
{|m|^3\over6\pi}-
 {2\sin(F\pi)\over3\pi^3 r^3}\, \int\limits_{|m|r}^\infty
dw(w^2-m^2r^2)^{3\over2}K_F(w)K_{1-F}(w)+$$
$$+{\sin(F\pi)\over \pi^3r^3}
\int\limits_{|m|r}^\infty dw\, w(w^2-m^2r^2)^{1\over2}
 {A\mu^{1-2F}({w\over r})^{2F}K_F^2(w)+A^{-1}
\mu^{2F-1}({w\over r})^{2(1-F)}K_{1-F}^2(w)\over A\mu^{1-2F}({w\over
r})^{2F}+2m+A^{-1}\mu^{2F-1}({w\over r})^{2(1-F)}}. \eqno(4.13)
$$

The effective potential in the massless theory is defined as
$$
{\cal U}^\eff(\x,\tau)=- {1\over2}\lim_{m\rightarrow 0}\tr
\langle\x,\tau|\ln \biggl[
{(-\partial_\tau^2+\tilde{H}^2)\tilde{m}^{-2}\over
(-\partial_\tau^2-\bpar^2+m^2)\tilde{m}^{-2}}\biggr]|\x,\tau\rangle.
\eqno(4.14)
$$
Defining the renormalized zeta function density
$$
\zeta_{\x,\tau}^\ren(z)=\lim_{m\rightarrow 0}
\bigl[\zeta_{\x,\tau}(z|m)- \zeta_{\x,\tau}^{(0)}(z|m)\bigr],
\eqno(4.15)
$$
we get, similarly to Eq.(4.9),
$$
{\cal U}^\eff(\x,\tau)={1\over2} \bigl[{d\over
dz}\zeta_{\x,\tau}^\ren(z)\bigr]\mid_{z=0}
+{1\over2}\zeta_{\x,\tau}^\ren(0)\ln\tilde{m}^2. \eqno(4.16)
$$
In the background of a singular magnetic vortex (1.1) -- (1.2), using
Eq.(4.10), we get
$$
\zeta_{\x,\tau}^\ren(z)=-{\sin(F\pi)\over 2\pi^{7\over2}}\cos(z\pi)
{\Gamma(z-{1\over2})\over\Gamma(z)} r^{2z-3}\times$$
$$\times
\left\{{\sqrt{\pi}\over4}\, {\Gamma({3\over2}-z)\over\Gamma(2-z)}
\left[1-{4F(1-F)\over 3-2z}\right] \Gamma\bigl({1\over2}-z+F\bigr)
\Gamma\bigl({3\over2}-z-F\bigr)+\right.$$
$$+\left.\int\limits_0^\infty dw\, w^{2(1-z)}
\bigl[K_F^2(w)-K_{1-F}^2(w)\bigr] \tanh\bigl[(2F-1)\ln({w\over\mu
r})+\ln A\bigr]\right\}, \eqno(4.17)
$$
in particular
$$\zeta_{\x,\tau}^\ren(0)=0, \eqno(4.18)
$$
and, thence, we arrive at the remarkable relation
$$
{\cal U}^\eff(\x,\tau)=\E_\x^\ren, \eqno(4.19)
$$
where $\E_\x^\ren$ is given by Eq.(4.2).

Although the last relation looks rather natural and even evident, let
us emphasize here that it is a consequence of the relation between the
renormalized zeta function densities of different spatial dimensions,
$$
\bigl[{d\over
dz}\zeta_{\x,\tau}^\ren(z)\bigr]\big|_{z=0}=-\zeta_\x^\ren(-{1\over2}),
\eqno(4.20)
$$
and relation (4.18). As it has been shown in Ref. \cite{SitR98},
relation (4.20) can be in general broken in spaces of higher
dimensions. Moreover, both left- and right-hand sides of Eq.(4.20) can
have nothing to do with the true vacuum energy density. Fortunately,
this is not relevant for the case considered in the present paper,
and, indeed, in the background of a singular magnetic vortex the vacuum
energy density coincides with the effective potential.

\section{Conclusion}

In the present paper we show that the massless fermionic vacuum under
the influence of a singular magnetic vortex (1.1) -- (1.2) in
2+1-dimensional space-time attains the energy
density (4.2) which decreases at large distances from the vortex as 
inverse power with integer exponent (4.5).

The most general set of boundary conditions at the location of the vortex
is used, see Eq.(2.14), providing the self-adjointness of the Dirac
Hamiltonian;  thus the vacuum energy density is depending on
self-adjoint extension parameter $\Theta$ or $A$ (3.19). As to the
dependence on vortex flux $\F$, it has been already anticipated in
Introduction that all vacuum polarization effects are gauge invariant
and thus depend on
reduced vortex flux $\F-\Upsilon$ rather than on $\F$
or $\Upsilon$ separately.  Note also that at half-integer values of $\F-\Upsilon$
(i.e. at $F={1\over2}$) the vacuum energy density is given by Eq.(4.3).

Among the whole variety of boundary conditions which are specified by
self-adjoint extension parameter $\Theta$, condition
$\cos\Theta=0$ (or $\Theta=\pm{\pi\over2}$) is distinguished, since it
corresponds to one of the two components of a solution to the Dirac
equation being regular for all $n$: if $\Theta=s{\pi\over2}$, then the
lower components are regular, and, if $\Theta=-s{\pi\over2}$, then the
upper components are regular. This condition
is parity invariant, and under it
the vacuum energy density is given by Eq.(4.4). It should be noted that this
condition is extensively discussed in the literature, being involved into
the two most popular ones: the condition of maximal simplicity
\cite{Alf}

$$\Theta=\left\{\ba{cc}
s{\pi\over2},& s(\F-\Upsilon)>0\\[0.2cm]
-s{\pi\over2},& s(\F-\Upsilon)<0\\ \ea \right\} \eqno(5.1)
$$
and the condition of minimal irregularity \cite{Sit90,Sit96}

$$\Theta=\left\{\ba{cc}
s{\pi\over2},& -{1\over2}<s(\fo\F-\Upsilon\fc-{1\over2})<0\\[0.2cm]
0,& \fo\F-\Upsilon\fc={1\over2}\\[0.2cm]
-s{\pi\over2},& 0<s(\fo\F-\Upsilon\fc-{1\over2})<{1\over2}\\ \ea
\right\}; \eqno(5.2)
$$
here, both in Eqs.(5.1) and (5.2), it is implied
that $\fo\F-\Upsilon\fc\neq0$.

Under condition (5.1) we get

$$
\E_\x^\ren=\left\{\ba{cc}
{\tan(\fo\F-\Upsilon\fc\pi)\over12\pi r^3}
\fo\F-\Upsilon\fc\bigl({1\over4}-\fo\F-\Upsilon\fc^2\bigr),&
\F-\Upsilon>0\\[0.2cm]
{\tan[(1-\fo\F-\Upsilon\fc)\pi]\over 12\pi
r^3}\bigl(1-\fo\F-\Upsilon\fc\bigr)
\bigl[{1\over4}-(1-\fo\F-\Upsilon\fc)^2\bigr],& \F-\Upsilon<0\\ \ea
\right\}.
\eqno(5.3) $$

Under condition (5.2) we get
$$
\E_\x^\ren=\left\{ \ba{cc}
{\tan(\fo\F-\Upsilon\fc\pi)\over24\pi r^3}
\bigl(\fo\F-\Upsilon\fc-{1\over2}\bigr)
\bigl[3|\fo\F-\Upsilon\fc-{1\over2}|-\\[0.2cm]
-2(\fo\F-\Upsilon\fc-{1\over2})^2-1\bigr],
& \fo\F-\Upsilon\fc\neq{1\over2}\\[0.2cm]
{1\over24\pi^2 r^3},& \fo\F-\Upsilon\fc={1\over2}\\ \ea
\right\}, \eqno(5.4)
$$

It is clear that Eq.(5.4), in contrast to Eq.(5.3), 
is periodic in the value of the vortex flux.

As it should be expected, the vacuum energy density is
invariant under transitions to equivalent representations of the
Clifford algebra (i.e. independent of $\chi_s$). It should be emphasized 
that the vacuum energy density is also invariant under the transition
to an inequivalent representation (i.e. under $s\rightarrow -s$).

\section*{Acknowledgements}

I am thankful to R.Jackiw, H.Leutwyler and W Thirring for interesting 
discussions. The research was supported by the State Foundation for Fundamental
Research of Ukraine.

\section*{Appendix}
\def\theequation{A.\arabic{equation}}
\setcounter{equation}{0}

Let us consider a general case of massive Hamiltonian $\tilde{H}$ (3.6). 
The relevant partial Hamiltonian has the form
\be
\tilde{h}_{n_0}=\left(\begin{array}{cc}m&e^{i\chi_s}[\partial_r+(1-F)r^{-1}]\\
e^{-i\chi_s}(-\partial_r-Fr^{-1})&-m\end{array}\right).
\ee
Let $\tilde{h}$ be the operator in the form of Eq.(A.1), which acts on
the domain of functions
$\xi^0(r)$ that are regular at $r=0$. Then its adjoint $\tilde{h}^{\dagger}$
which is defined by the relation
\be
\int^\infty_0 dr\,r[\tilde{h}^\dagger\xi(r)]^\dagger \xi^0(r) =
\int^\infty_0 dr\,r[\xi(r)]^\dagger[\tilde{h}\xi^0(r)]
\ee
acts on the domain of functions $\xi(r)$ that are not necessarily
regular at $r=0$. So the question is, whether the domain of definition
of $\tilde{h}$ can be extended, resulting in both the operator and its
adjoint being defined on the same domain of functions? To answer this,
one has to construct the eigenspaces of $\tilde{h}^\dagger$
with complex eigenvalues. They are spanned by the linearly independent
square-integrable solutions correspoding to the pair of purely imaginary
eigenvalues,
\be
\tilde{h}^\dagger \xi^\pm (r) = \pm i\mu\xi^\pm (r),
\ee
where $\mu>0$ is inserted for the dimension reasons. It can be shown that in the
case of Eq.(A.1) only one pair of such solutions exists, thus the deficiency
index of $\tilde{h}$ is equal to (1,1). This pair is given by the following
expression
\be
\xi^\pm (r) = {1\over N}\left(\begin{array}{l}e^{i\chi_s} \exp{\bigl[\pm{i\over2}
\sgn (m) \eta\bigr]}K_F(\tilde{\mu} r)\\
\sgn (m) \exp{\bigl[\mp{i\over2}\sgn (m) \eta\bigr]}K_{1-F}(\tilde{\mu}
r)\end{array}\right),
\ee
where $N$ is a certain normalization factor and
\be
\tilde{\mu} = \sqrt{\mu^2 + m^2},\,\,  \eta = \arctan\biggl({\mu\over|m|}\biggr).
\ee
Self-adjoint extended operator $\tilde{h}^{\theta_s}$ is defined on
the domain of functions of the form
\be
\left(\ba{c}
\tilde{f}_{n_0}\\
\tilde{g}_{n_0}\\ \ea
\right) = \xi^0 + c(\xi^+ + e^{i\theta_s}\xi^-),
\ee
where $c$ is a complex parameter and $\theta_s$ is a real continuous
parameter which depends, in general, on the choice between the two
inequivalent representations of the Clifford algebra. Using the
asymptotics of the Macdonald function at small values of the variable,
we get
\be
\left(\begin{array}{c}\tilde{f}_{n_0}\\ \tilde{g}_{n_0}\end{array}\right)
\mathop{\sim}_{r\to0}
\left(\begin{array}{l}e^{i\chi_s} \cos\left\{{1\over2}[\theta_s-\sgn (m)
\eta]\right\}
2^F\Gamma (F)(\tilde{\mu} r)^{-F}\\
\sgn (m)\cos\left\{{1\over2}[\theta_s+\sgn (m)\eta]\right\}2^{1-F}\Gamma
(1-F)(\tilde{\mu}r)^{-1+F}\end{array}\right),
\ee
or
\bea
\left\{\tan\big[{1\over2}\theta_s - {1\over2}\sgn (m) \eta\big] \sin \eta
- \sgn (m)\cos\eta\right\} \lim_{r\to 0}(\tilde{\mu}r)^F\tilde{f}_{n_0}=
\nonumber
\eea
\bea=
-e^{i\chi_s}2^{2F-1}{\Gamma(F)\over\Gamma(1-F)}\lim_{r\to 0}(\tilde\mu
r)^{1-F}\tilde{g}_{n_0}.
\eea
Defining new parameter $\Theta$ by means of relation
\bea
&& \tan\biggl(s{\Theta\over2} +{\pi\over4}\biggr)=\left\{\tan\big[{1\over2}
\theta_s-{1\over2}\sgn(m)\eta\big]\sin\eta-\sgn(m)\cos\eta\right\}^{-1}
 \biggl({2\mu\over\tilde{\mu}}\biggr)^{2F-1}{\Gamma(F)\over\Gamma(1-F)},
\eea
we get Eq.(2.14).

Certainly, both $\theta_s$ and $\Theta$ can be regarded as self-adjoint
extension parameters which are to specify the boundary condition at $r=0$. The
use of $\Theta$ in this aspect may seem to be more preferable just for
the convenience reasons, because Eq.(2.14) looks much simpler than
Eq.(A.8). In particular, Eq.(2.14), in contrast to Eq.(A.8), is independent
of $m$ and remains explicitly invariant under $s\to-s$ ($\Theta$ is
independent of $s$).

In the limit of $m\to \pm 0$ Eq.(A.9) takes the form
\be
\tan\biggl(s{\Theta\over2}+{\pi\over4}\biggr)=-\tan\big[{1\over2}
\theta_s+{1\over4}\sgn(m)\pi\big]2^{2F-1}{\Gamma(F)\over\Gamma(1-F)},
\ee
then we get
\be
\theta_s=s\theta,
\ee
where $\theta$ is independent of s.

Concluding this appendix, let us note that all characteristics of the
massless fermionic vacuum are depending on $A$ (3.19) rather than on
$\Theta$ itself, and $A$ is expressed through $\theta$ in the following way
\be
A=-\tan\big[{1\over2}s\theta + {1\over4}\sgn(m) \pi\big].
\ee

\end{document}